\documentclass[aps,prd,nofootinbib,showkeys,preprint,floatfix]{revtex4}

\usepackage[utf8]{inputenc}
\usepackage{amssymb}
\usepackage{amsmath}
\usepackage{amsfonts}
\usepackage{graphicx}
\usepackage{color}
\usepackage{xspace}
\usepackage{ulem}
\usepackage{lscape}
\usepackage{ wasysym }

\usepackage{multirow,rotating}

\def\gsim{\raise0.3ex\hbox{$\;>$\kern-0.75em\raise-1.1ex\hbox{$\sim\;$}}}
\def\lsim{\raise0.3ex\hbox{$\;<$\kern-0.75em\raise-1.1ex\hbox{$\sim\;$}}}

\def\znbb{0\nu\beta\beta}
\def\meff{\langle m_{\nu} \rangle}

\newcommand{\ba}[1]{\begin{eqnarray} \label{(#1)}}
\newcommand{\ea}{\end{eqnarray}}

\newcommand{\AddrAHEP}{
  {\it AHEP Group, Instituto de F\'{\i}sica Corpuscular --
    C.S.I.C./Universitat de Val{\`e}ncia \\
    Edificio de Institutos de Paterna, Apartado 22085,
  E--46071 Val{\`e}ncia, Spain}}

\newcommand{\AddrUFSM}{
Universidad T\'ecnica Federico Santa Mar\'\i a, \\ 
Centro-Cient\'\i fico-Tecnol\'{o}gico de Valpara\'\i so, \\ 
Casilla 110-V, Valpara\'\i so,  Chile}

\def\gsim{\raise0.3ex\hbox{$\;>$\kern-0.75em\raise-1.1ex\hbox{$\sim\;$}}}
\def\lsim{\raise0.3ex\hbox{$\;<$\kern-0.75em\raise-1.1ex\hbox{$\sim\;$}}}

\begin{document}

\preprint{IFIC/16-41}  

\title{Scalar-mediated double beta decay and LHC}

\author{L. Gonzales}\email{lorena.gonzalez@alumnos.usm.cl}\affiliation{\AddrUFSM}
\author{J.C. Helo} \email{juan.heloherrera@gmail.com}\affiliation{\AddrUFSM}
\author{M. Hirsch} \email{mahirsch@ific.uv.es}\affiliation{\AddrAHEP}
\author{S.G. Kovalenko}\email{Sergey.Kovalenko@usm.cl}\affiliation{\AddrUFSM}



\begin{abstract}
The decay rate of neutrinoless double beta ($0\nu\beta\beta$) decay
could be dominated by short-range diagrams involving heavy scalar
particles (``topology-II'' diagrams). Examples are diagrams with
diquarks, leptoquarks or charged scalars.  Here, we compare the
discovery potential for lepton number violating signals at the LHC
with constraints from dijet and leptoquark searches and the
sensitivity of $\znbb$ decay experiments, using three example
models. We note that already with 20/fb the LHC will test interesting
parts of the parameter space of these models, not excluded by current
limits on double beta decay. 

\end{abstract}

\maketitle
\section{Introduction}
\label{sect:intro}

From the theoretical point of view, neutrinoless double beta decay
($\znbb$) can be written as a dimension-9 operator:
\begin{equation}\label{eq:d9}
{\cal O}^{\znbb} = \frac{c_9}{\Lambda_{LNV}^5} 
                 {\bar u}{\bar u}dd {\bar e} {\bar e}.
\end{equation}
Here, $\Lambda_{LNV}$ is the scale of lepton number violation (LNV).
Many beyond the standard model contributions to this operator have
been discussed in the literature, for a review see
\cite{Deppisch:2012nb}.  Contributions to the decay rate of $\znbb$
decay can be classified as (i) neutrino mass mechanism; (ii)
long-range \cite{Pas:1999fc} and (iii) short-range contributions
\cite{Pas:2000vn}. \footnote{Neither in the long-range nor the
  short-range part of the amplitude the neutrino mass does appear
  directly. However, the $\Delta L=2$ interactions, present
  necessarily in all contributions to $\znbb$ decay, implies Majorana
  neutrino masses must be non-zero in all possible models contributing
  to eq.(\ref{eq:d9}) \cite{Helo:2015fba,Helo:2016vsi}.}

Particularly interesting is the possibility that all
beyond-standard-model particles, appearing in the ultra-violet
completions of this operator, are heavy. This corresponds to the
short-range part of the $\znbb$ decay amplitude. In this case, with
the current sensitivities of $\znbb$ decay experiments
\cite{Agostini:2013mzu,KamLAND-Zen:2016pfg} of the order of roughly
${\cal O}(10^{25}-10^{26})$ yr, one probes mass scales in the range
$\Lambda_{LNV} \sim (1-3)$ TeV - exactly the range of energy explored
at the LHC.

\begin{figure}[t] 
\hskip10mm\includegraphics[width=0.7\linewidth]{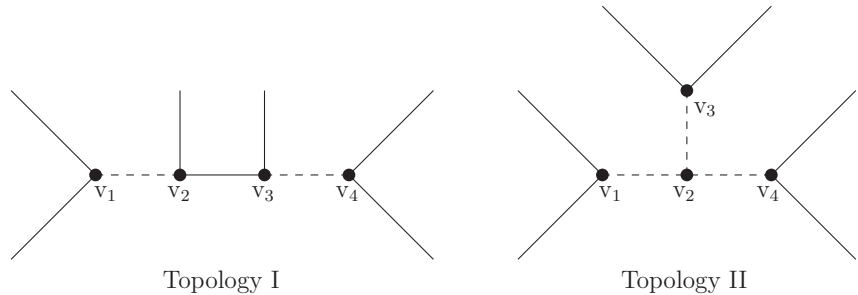}
\caption{\it \label{Fig:0nbbTopologies}Tree-level topologies for the
  $d=9$ $0\nu\beta\beta$ decay operator. External lines are fermions;
  internal lines can be fermions (solid) or scalars/vectors.  For T-II
  there are in total 4 possibilities classified as: SSS, VVV, SSV and
  VVS. Only SSS and VVS can contribute significantly to $\znbb$ decay
  \cite{Bonnet:2012kh}. We will concentrate on scalar-only
  contributions.}
\end{figure}

A list of all possible decompositions of eq. (\ref{eq:d9}) has been
found in \cite{Bonnet:2012kh}. Models fall into two classes, called
topology-I (T-I) and topology-II (T-II), see
fig. (\ref{Fig:0nbbTopologies}). In this figures outside lines
correspond to the six fermions appearing in eq. (\ref{eq:d9}) , while
the internal particles can be scalars, vectors or fermions.
Just to mention one example for T-I and T-II each: In left-right (LR)
symmetric models, right-handed gauge bosons ($W_R$) and neutrinos
($N_R$) appear in T-I as $W_R-N_R-W_R$ exchange
\cite{Mohapatra:1980yp,Mohapatra:1986pj}, while a T-II type diagram 
can appear as $W_R-\Delta^{\pm\pm}_R-W_R$ exchange \cite{Hirsch:1996qw}
in LR models with right-handed triplets ($\Delta_R$). 

The classical LNV signal searched for at the LHC is two same-sign
leptons plus jets ($lljj$), first discussed as a possible signal for
left-right symmetric models in \cite{Keung:1983uu}, see also
\cite{Das:2012ii}. This signal is generated from the T-I diagram with
right-handed neutrinos. The doubly charged scalar can be
searched via vector-boson-fusion, see for example
\cite{Bambhaniya:2015wna,Dev:2016dja}. This corresponds to the T-II
diagram mentioned above. VBF gives the same final state (lljj), but
has different kinematics.  We mention in passing that also di-lepton
searches can be used to put bounds on LR models
\cite{Lindner:2016lpp}.

Both ATLAS and CMS have published results for run-I of the LHC.  CMS
\cite{Khachatryan:2014dka} observed an excess in the electron sample
around $m_{eejj} \simeq 2$ TeV,\footnote{14 events with an estimated
  background of 4 events \cite{Khachatryan:2014dka}, roughly equal to
  $2.8$ $\sigma$ c.l.} but no excess in the muon sample. CMS
interprets the excess as a statistical fluctuation. ATLAS used
$20.3$/fb of pp collision data in their search \cite{Aad:2015xaa},
finding no anomalous events. The experimental collaborations then give
limits on heavy Majorana neutrinos in left-right (LR) symmetric
models, derived from this data.

However, LNV searches at the LHC do not give bounds only for LR
models. In principle, all models that contribute to eq.(\ref{eq:d9})
via short-range contributions should lead to a LNV signal at the LHC.
For the case of topology-I, the implications of LNV searches at the
LHC and their connection to $\znbb$ decay has been studied in
\cite{Helo:2013dla,Helo:2013ika}.  In this paper we will study future
LHC constraints on topology-II models.  We will concentrate on the
case where the non-SM particles are all scalars.

Both, ATLAS and CMS have published searches using dijets, based on
$\sqrt{s}=8$ TeV \cite{Aad:2014aqa,Khachatryan:2015sja} and
$\sqrt{s}=13$ TeV \cite{ATLAS:2015nsi,Khachatryan:2015dcf} data.  No
new resonances have been observed in these searches, both
collaborations give instead upper limits on $\sigma\times BR$ as a
function of resonance mass.  While dijet data of course can not be
used to establish the existence of LNV, non-observation of new
resonances in dijet searches at the LHC can be used to obtain
limits on $\znbb$ decay \cite{Helo:2015ffa}.  In our analysis,
presented below, we will also estimate the reach of future LHC data
and compare it to expectations for the LNV searches.

As discussed below, in many of the models for T-II double beta decay
leptoquarks (LQs) appear. Searches for leptoquarks have been carried
out at the LHC by both ATLAS and CMS. Lower limits on the masses of
first generation LQs from pair production in the $\sqrt{s}=8$ TeV data
are now roughly of the order of 1 TeV
\cite{Khachatryan:2015vaa,Aad:2015caa}. ATLAS has published first
limits from $\sqrt{s}=13$ TeV data with only $3.2/$fb, which already
give very similar limits \cite{Aaboud:2016qeg} despite the smaller
statistics. Searches for singly produced LQs, published by CMS
\cite{Khachatryan:2015qda}, give more stringent limits, albeit only
for large values of the LQ coupling to quarks and leptons. Also 
these limits and results of future searches can be used to constrain 
short-range contributions to double beta decay and we take into 
account these constraints in our numerical analysis.

The rest of this paper is organized as follows. In section
\ref{sec:decomp} we discuss different T-II contributions to $\znbb$
decay. We give the Lagrangian and necessary definitions for three
example models. These models cover the optimistic/pessimistic cases
for $\znbb$ decay. In section \ref{sect:pheno}, we present our
numerical results. We then close with a short summary and discussion.

\section{General setup}
\label{sec:decomp}

In this section we will first recall the general setup of the
topology-II contributions to $\znbb$ decay. We will then give a few
more details for those three concrete example models, that we will
study numerically in section \ref{sect:pheno}. These examples, 
chosen from the full list of possible scalar models given in
\cite{Bonnet:2012kh}, allow us to cover both the most optimistic 
and the most pessimistic cases for the sensitivity of future 
double beta decay experiments. 

\subsection{Topology-II decompositions}

\begin{table}[t]
\begin{center}
\begin{tabular}{ccccl}
\hline \hline 
&& 
\multicolumn{3}{c}{Mediator $(Q_{\rm em}, Q_{\text{colour}})$} \\
\# & Decomposition & $S$  & $S'$  & $S'' $   \\
\hline 
1 
&
$(\bar{u} d) (\bar{u} d) (\bar{e} \bar{e})$
&
 (+1, {\bf 1} or {\bf 8})
&
 (+1, {\bf 1} or  {\bf 8})
&
$ (-2, {\bf 1})$
\\
2
&
$(\bar{u} d) (\bar{u} \bar{e}) (\bar{e} d)$
&
 (+1, {\bf 1} or  {\bf 8})
&
$(-1/3,{\bf 3})$
&
$(-2/3,\overline{\bf 3})$
\\
3
&
$(\bar{u} \bar{u}) (dd) (\bar{e} \bar{e})$
&
(+4/3, $\overline{\bf 3}$ or  {\bf 6})
&
(+2/3, {\bf 3} or  $\overline{\bf{6}}$)
&
$(-2,{\bf 1})$
\\
4
&
$(\bar{u} \bar{u}) (\bar{e} d) (\bar{e} d)$
&
 (+4/3, $\overline{\bf 3}$ or  {\bf 6})
&
$ (-2/3, \overline{\bf 3})$
&
$ (-2/3, \overline{\bf 3})$
\\
5
&
$(\bar{u} \bar{e}) (\bar{u} \bar{e}) (d d)$
&
$ (-1/3,{\bf 3})$
&
$ (-1/3,{\bf 3})$
&
 (+2/3, {\bf 3} or $\overline{\bf{6}}$)
\\
\hline \hline
\end{tabular}
\end{center}
\caption{\it \label{Tab:TopoII} List of decompositions for topology~II
  from \cite{Bonnet:2012kh}.  Only the electric and colour charges of
  the internal bosons are given here.  All listed possibilities give
  short-range contributions. For the colour charges in some cases
  there exist two possible assignments.}
\end{table}

Considering only the unbroken $SU(3)_C$ and $U(1)_{Q}$ there are only
five possible decomposition of eq. (\ref{eq:d9}) for topology-II.
These are listed in table \ref{Tab:TopoII}. Note that in some cases
there is more than one possibility for colour.  There are six scalar
states in these decompositions: (i) charged scalars, $S^{+}$ and $S^{--}$;
(ii) diquarks, $S^{4/3}_{DQ}$ and $S^{2/3}_{DQ}$; and (iii) leptoquarks,
$S^{-2/3}_{LQ}$ and $S^{-1/3}_{LQ}$.

Depending on the chirality of the outer fermions, the diquarks could
come either from electro-weak (EW) singlets or triplets, while the
leptoquarks could either be members of singlets or doublets. We have 
examples for each in the three selected models below. The singly 
charged scalar $S^{+}$ necessarily has to be a member of an 
$SU(2)_L$ doublet: $S_{1,2,1/2}$. Here and everywhere else in this 
paper the subscripts give the transformation properties under 
the SM group in the order $SU(3)_C\times SU(2)_L\times U(1)_Y$. 
Finally, $S^{--}$ could either come from an EW singlet or a triplet. 

Considering the full SM group, overall \cite{Bonnet:2012kh} gives 27
different combinations (``models'') for the five decompositions shown
in table \ref{Tab:TopoII}. All of these generate Majorana neutrino
masses, from tree-level masses for decompositions with $S_{1,3,-1}$ to
4-loop neutrino masses for the diagram containing
$S_{3,1,-1/3}-S_{3,1,-1/3}-S_{{\bar 6},1,2/3}$ \cite{Helo:2015fba}. 
Our three examples correspond to two 2-loop and
one 1-loop model, see below. This is motivated by the fact that for
2-loop neutrino mass models one can expect that the short-range part
of the amplitude for $\znbb$ and the mass mechanism can give similar
contributions to the overall decay rate \cite{Helo:2015fba}.

\subsection{Selected example models}

\begin{figure}
\centering
\includegraphics[scale=0.55]{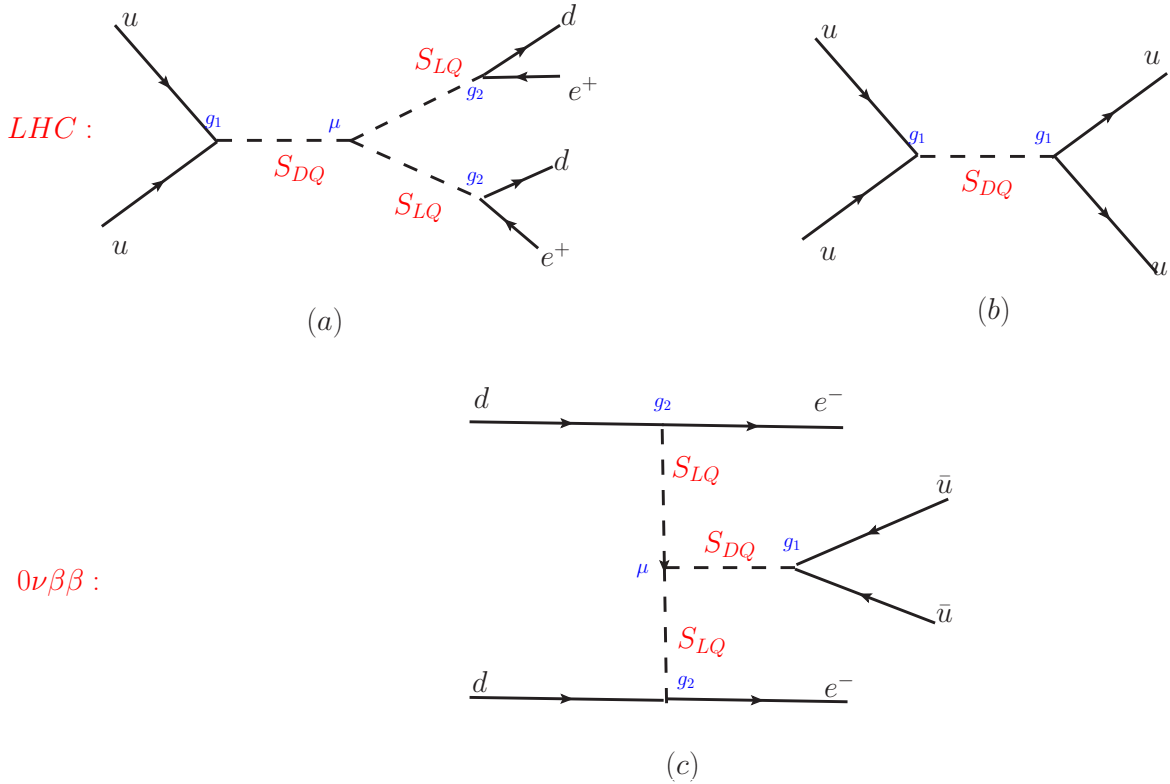}
\caption{\label{fig:DQM} Quark-level Feynman diagrams for 
(a) same-sign dilepton plus jets ($lljj$) signal, (b) dijet 
signal at the LHC and (c) neutrinoless double beta decay for 
the example model-1 containing a diquark and a leptquark scalar 
state.}
\end{figure}

Here, we will give the basic Lagrangian terms of three decompositions
of the $d=9$ $\znbb$ decay operator taken from \cite{Bonnet:2012kh}. 
These examples correspond to T-II-2 BL $\#$ 11, T-II-4 BL $\#$ 11 and 
T-II-5 BL $\#$ 11 in the notation of \cite{Bonnet:2012kh}.  Constraints 
on other short-range T-II decompositions will be very similar to 
these examples, as we will also discuss in section \ref{sect:pheno}.

\subsubsection{T-II-4, BL $\#$ 11}
\label{M1}

Our first example model contains two new particles: A scalar diquarks
and a leptoquark. In the context of $\znbb$ decay, diquark
contributions were first discussed in \cite{Gu:2011ak}. We define
scalar diquarks as particles coupling to a pair of same-type
quarks. We choose the example T-II-4, BL $\#$ 11 in the notation of
\cite{Bonnet:2012kh}.  This model generates neutrino masses at 2-loop
order \cite{Helo:2015fba}, which means the TeV scale is the natural
scale to fit to neutrino data. One expects therefore that this model
is testable at the LHC. Note that a possible SU(5) embedding of this
model has been recently discussed in \cite{Fonseca:2015ena}.

The new beyond the SM states in this model are:
\begin{eqnarray}\label{eq:2-mod-1}
\nonumber
&& S_{DQ} = S_{6,3,1/3} = \begin{pmatrix} S^{1/3}_{DQ} & S^{4/3}_{DQ} \\[3mm] S^{-2/3}_{DQ} &
-S^{1/3}_{DQ} \end{pmatrix},  \ \ \ \  
S_{LQ} = S_{3,2,1/6} = \begin{pmatrix} S^{2/3}_{LQ}  \\[3mm] S^{-1/3}_{LQ}\end{pmatrix}.
\end{eqnarray}

%
%
The interaction Lagrangian of the model is given by:
 \begin{eqnarray}\label{eq:3-modif-1}
%
&&{\cal L}^{(1)}_{DQLQ} = {\cal L}_{SM} + g_{1}\  \bar{Q} \tau_{2} \cdot\hat{S}_{DQ} \cdot  Q^{c} +
g_{2}\  \bar{L} \tau_{2} \cdot S^{\dagger}_{LQ} \cdot d_{R} + \mu \ S^{\dagger}_{LQ} \tau_{2}\cdot
\hat{S}_{DQ} \cdot S^{\dagger}_{LQ} + {\rm h.c.}
\end{eqnarray}
Here we introduced the notation $\hat{S}_{DQ}= S_{DQ, a}^{(6)}
(T_{\bar{\bf 6}})^{a}_{IJ}$, with $I,J=1-3$ and the color triplet
indexes and $a=1-6$ the color sextet indexes.  $g_1$ and $g_2$ are
dimensionless Yukawas and $\mu$ has dimension of mass. The symmetric
$3\times 3$ matrices $T_{\bf 6}$ and $T_{\bar{\bf 6}}$ can be found in
ref.~\cite{Bonnet:2012kh}. Note that eq. (\ref{eq:3-modif-1}) violates 
lepton number by two units. 

The inverse half-life for $\znbb$ for the diagram of figure
\ref{fig:DQM}, is given by \cite{Bonnet:2012kh}:
\begin{eqnarray}\label{NLDBD-DQ}
T^{-1}_{1/2} =  G_{01} \left|\epsilon_{DQ} {\cal{M}}_{DQ} \right|^2,
\end{eqnarray} 
where $G_{01}$ is a phase space integral and $\epsilon_{DQ}$ is
defined by
\begin{equation}\label{eq:eps}
\epsilon_{DQ} = \frac{2 m_p}{G_F^2}\frac{g_1g_2^2\mu}{m_{DQ}^2m_{LQ}^4},
\end{equation}
and the nuclear matrix element is:
\begin{equation}\label{eq:MDQ}
{\cal{M}}_{DQ} = \frac{1}{48} {\cal{M}}_{1} -  \frac{1}{192}  {\cal{M}}_{2}.
\end{equation}
Here ${\cal{M}}_{1,2}$ are defined in  \cite{Pas:2000vn}, 
numerical values for $^{136}$Xe can be found in \cite{Deppisch:2012nb}.

\subsubsection{T-II-5, BL $\#$ 11}
\label{M2}

As a second example we discuss another model with a scalar diquark.
However, this diquark couples only to down-type quarks. This model was
first discussed in \cite{Kohda:2012sr}. It corresponds to the example
T-II-5, BL$\#$ 11 from the list of decompositions of the $d=9$ $\znbb$
decay operator \cite{Bonnet:2012kh}. Also this model generates neutrino
masses at 2-loop order as discussed in \cite{Helo:2015fba}.

This particular case introduces a singlet diquark $S^{2/3}_{DQ} = S_{\bar{6}, 1,2/3}$ and a singlet leptoquark
 $S^{1/3}_{LQ} = S_{\bar{3}, 1, 1/3}$. With these new fields, the Lagrangian contains the interactions:
\begin{eqnarray}\label{eq:7-modif-1}
\nonumber
&&{\cal L}^{(2)}_{DQLQ} = {\cal L}_{SM} + g_{1}\  \bar{d^{c}_{R}}\cdot\hat{S}^{2/3}_{DQ} \cdot  d_{R} +
g_{2}\  \bar{L} \tau_{2} \cdot Q^{c}\cdot   S^{1/3\, \dagger}_{LQ}  + \mu \ S^{1/3\, \dagger}_{LQ} \cdot
\hat{S}^{2/3}_{DQ} \cdot S^{1/3\, \dagger}_{LQ} + {\rm h.c.}
\end{eqnarray}
Here, as before, by definition $\hat{S}^{2/3}_{DQ} = S^{2/3}_{DQ, a} (T_{\bf {6}})^{a}_{IJ}$.
%

The inverse half-life for the short-range $\znbb$ decay in this model
has the same form as eq. (\ref{NLDBD-DQ}) (with some obvious
replacements). In particular, it depends in the same combination of
nuclear matrix elements.

\subsubsection{T-II-2, BL$\#$
11 }
\label{M3}
Finally, we will discuss a model with a singly charged scalar.  We
choose the example T-II-2, BL$\#$11 from the list of
\cite{Bonnet:2012kh}.  This model generates neutrino masses at 1-loop
order \cite{Helo:2015fba}.

In this model, we add the following states to the SM particle content:  
\begin{align}
S_{1,2,1/2}
 =&
 \begin{pmatrix}
 S^{1}
  \\
  S^{0}
 \end{pmatrix}, \hskip 10mm
 S_{LQ}=S_{3,2,1/6}
  =
 \begin{pmatrix}
  S_{LQ}^{(2/3)}
  \\
  S^{(-1/3)}_{LQ}
 \end{pmatrix},
 \hskip 10mm
S^{1/3}_{LQ}=S_{\bar{3},1,1/3}.
\end{align}
With these new fields, the relevant Lagrangian is:
 %
%
 %
 \begin{eqnarray}\label{eq:9-modif-1}
\nonumber
{\cal L}_{S_{1} LQ} &=& {\cal L}_{SM} + g_{1}\  \bar{Q}\cdot S_{1,2,1/2} \cdot  d_{R} +
g_{2}\  \bar{Q} \tau_{2} \cdot L^{c}\cdot   S^{1/3\, \dagger}_{LQ}  
+ g_{3}\  \overline{d_{R}} \ L \tau_{2} \cdot  S_{LQ}  \\   \nonumber && +  \ \mu \ S^{\dagger}_{1,2,1/2} \cdot S_{LQ} \cdot S^{1/3}_{LQ} + {\rm h.c.}
\end{eqnarray}
The inverse half-life for $\znbb$ (short-range part of the amplitude) 
can be written as:
\begin{eqnarray}\label{NLDBD-SC}
T^{-1}_{1/2} =  G_{01} \left|\epsilon_{S_1} {\cal{M}}_{S_1} \right|^2,
\end{eqnarray} 
where $\epsilon_{S_1}$ is given by
\begin{equation}\label{eq:epsS1}
\epsilon_{S_1} = \frac{2 m_p}{G_F^2}\frac{g_1g_2g_3\mu}{m_{S_1}^2m_{LQ}^4},
\end{equation}
and the matrix element is given by:
\begin{equation}\label{eq:MS1}
{\cal{M}}_{S_1} = -\frac{1}{16} {\cal{M}}_{1}.
\end{equation}
Again, for further definitions and numerical values see
\cite{Pas:2000vn,Deppisch:2012nb}.

 \section{Numerical results}
\label{sect:pheno}

In this section we present our numerical results.  We estimate the
sensitivity of current and future $\znbb$ experiments and compare them
with the sensitivity of dijet, leptoquark and dilepton plus jets
searches at LHC at $\sqrt{s} = 13 \ \text{TeV}$. For definiteness 
we assume two values for the accumulated luminosity ${\cal L}$: 
${\cal L}= 20/$fb and  ${\cal L}= 300/$fb.

For the calculation of the cross sections of the diquark scalar
resonances we use MadGraph5 \cite{Alwall:2014hca}, for the leptoquark
and the singly charged scalar CalcHEP \cite{Pukhov:2004ca}.  We have
compared our results with the literature \cite{Han:2010rf} and found
good agreement with published values, whenever available. Plots for
the cross sections can be found in our previous work on T-I contributions 
for $\znbb$ decay \cite{Helo:2013ika}.

From the cross sections we then estimate the future LHC sensitivity as
follows. For the LNV signal (lljj) we first take a simple fit
\cite{Helo:2013ika} to the background of existing data of the CMS
analysis \cite{CMS:2012uwa} based on 3.6 $\text{fb}^{-1}$ at $\sqrt{s}
= 8$ TeV. We checked this fit against the CMS analysis
\cite{Khachatryan:2014dka} based on 19.7 $\text{fb}^{-1}$ of data at
$\sqrt{s} = 8$ TeV, published later, and found good overall
agreement. In the CMS analysis \cite{Khachatryan:2014dka} the main
background can be traced to $t \bar{t}$ events. We then do a simple
estimate which considers that the $t \bar{t}$ production cross section
is very roughly about a factor 3 higher at $\sqrt{s} = 13
\ \text{TeV}$ than at $\sqrt{s} = 8 \ \text{TeV}$. Thus, we scale the
original fit to $\sqrt{s} = 8$ TeV data with a simple constant and
scale the background function from ${\cal L} = 3.6 \ \text{fb}^{-1}$
to future expected luminosities of ${\cal L} = 20/$fb and
$300/$fb. For the estimation of the future dijet background we use the
fit of the SM dijet distribution fitted to Monte Carlo simulation
given in \cite{Richardson:2011df}.  For both, dijet and $lljj$
analysis we then estimate backgrounds as dicussed above and define the
sensitivity reach as either the simple square root of the background
(times two for 95 \% c.l.) or 5 signal events, whichever is
larger. For future LQ searches at the LHC, we calculate LQ pair
production cross sections as a function of LQ mass. We simply define
the reach of the LQ search then as the mass for which there are less
then 10 signal events in $20/$fb ($300/$fb) at the LHC (before
cuts). This results in the simple estimate of $m_{LQ} \gsim 1.3$ TeV
($m_{LQ} \gsim 1.8$ TeV) as the near (far) future limit.  Thus, our
results should be considered only rough estimates. For more exact
results a full MonteCarlo simulation including detector effects would
be necessary.

For double beta decay we use the current limit of
$T_{1/2}^{\znbb}(^{136}Xe) \ge 1.1 \times 10^{26}$ yr from the
KamLAND-Zen collaboration \cite{KamLAND-Zen:2016pfg}.\footnote{For the
  mass mechanism this limit corresponds to $\meff \lsim 0.1$ ($0.14$)
  eV, depending on nuclear matrix elements \cite{Simkovic:2013qiy}
  (\cite{Muto:1989cd}).}  Several experimental proposals aim at
half-life sensitivities of the order of $10^{27}$ yr. We will use the
estimated sensitivity of the nEXO proposal
\cite{Pocar:2015mrz,Pocar:2015ota} of $T_{1/2}^{\znbb}(^{136}Xe)
\simeq 6 \times 10^{27}$ yr for our calculation of the future limits.
We convert half-life limits into limits on masses and couplings, 
using the equations discussed in the previous section. We take into 
account the QCD corrections to the Wilson coefficients, calculated 
recently in \cite{Gonzalez:2015ady}. In particular for the model 
with the singly charged scalar QCD corrections have been found to 
be very important numerically. 

We will first discuss the case of our example model 1, see the
Lagrangian in eq. (\ref{eq:3-modif-1}). In this model the three components
of the triplet diquark, the scalars $S^{(4/3)}_{DQ}, S^{(1/3)}_{DQ},
S^{(-2/3)}_{DQ}$, contribute to the dijet cross section. However, the
dominant contribution to the dijet cross section comes from the
diquark scalar $S^{(4/3)}_{DQ}$. The Feynman diagram is shown in
fig. \ref{fig:DQM}. We have assumed for simplicity that the Yukawa
couplings $g_1$ and $g_2$ are different from zero for the first quark
and lepton generations only.  As is shown in fig. \ref{fig:DQM}, the
scalar diquark $S^{(4/3)}_{DQ}$ can only decay through two possible
channels: dijets (jj) and dilepton plus two jets (lljj). The
respective branching ratios can be calculated directly from the
Lagrangian (\ref{eq:3-modif-1}) and are a function of the leptoquark mass
$m_{LQ}$ and the (unknown) parameters $\mu$ and $g_2$.

\begin{figure}
\centering
\includegraphics[scale=0.6]{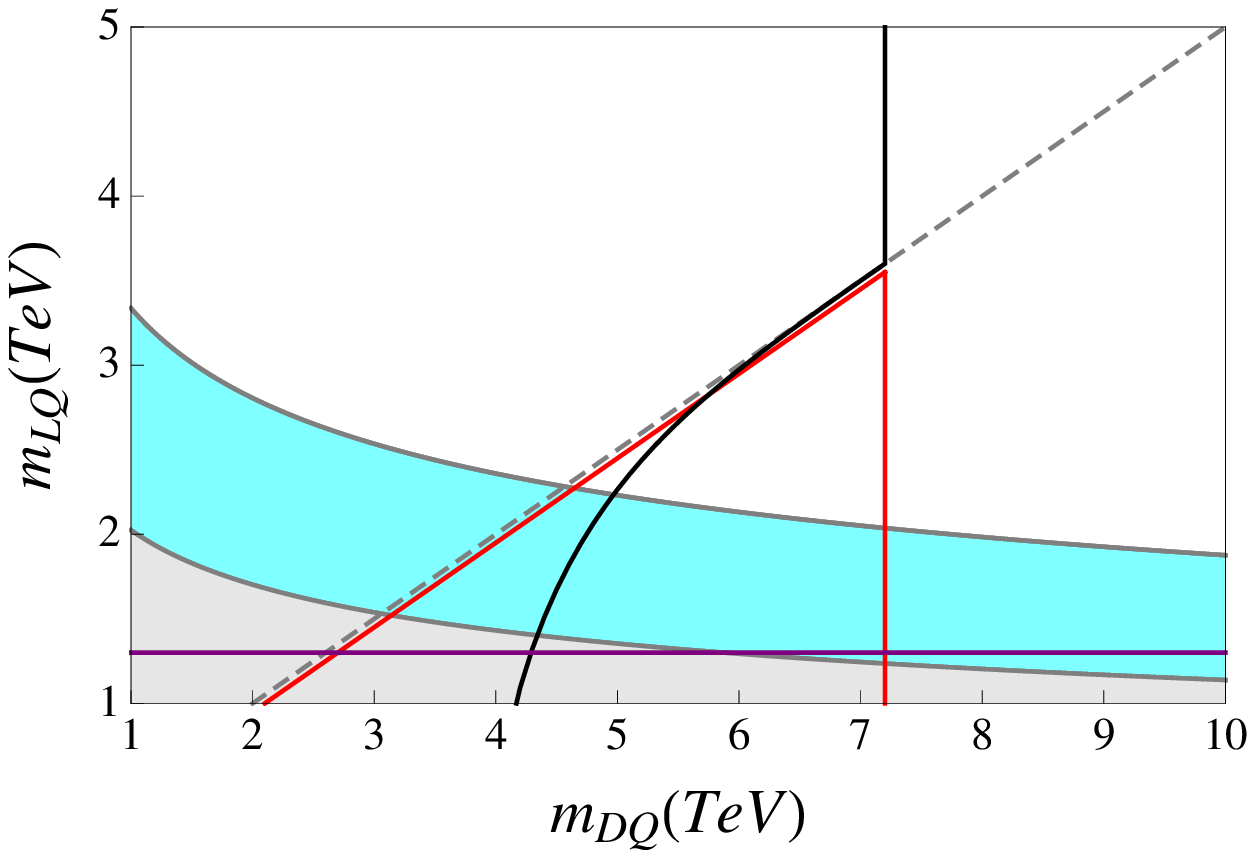}
\includegraphics[scale=0.6]{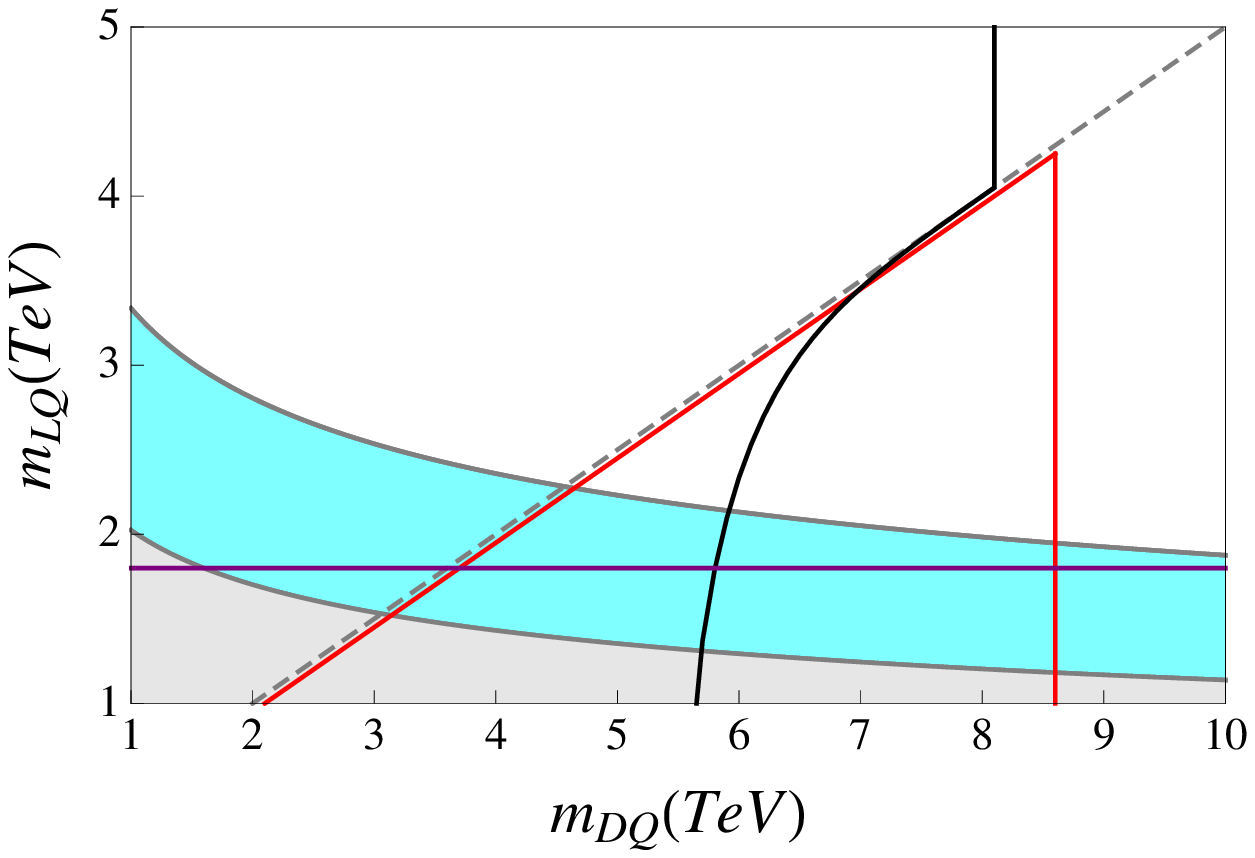}
\includegraphics[scale=0.6]{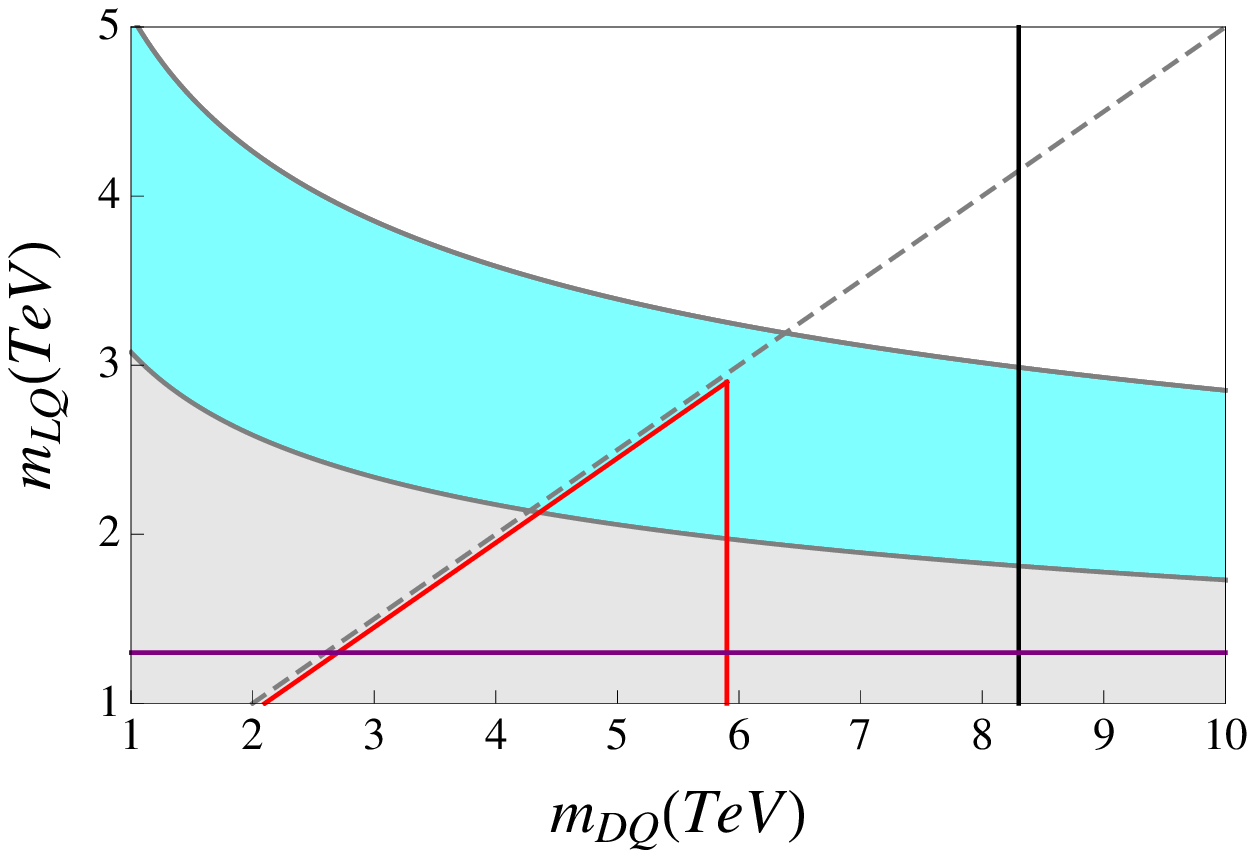}
\includegraphics[scale=0.6]{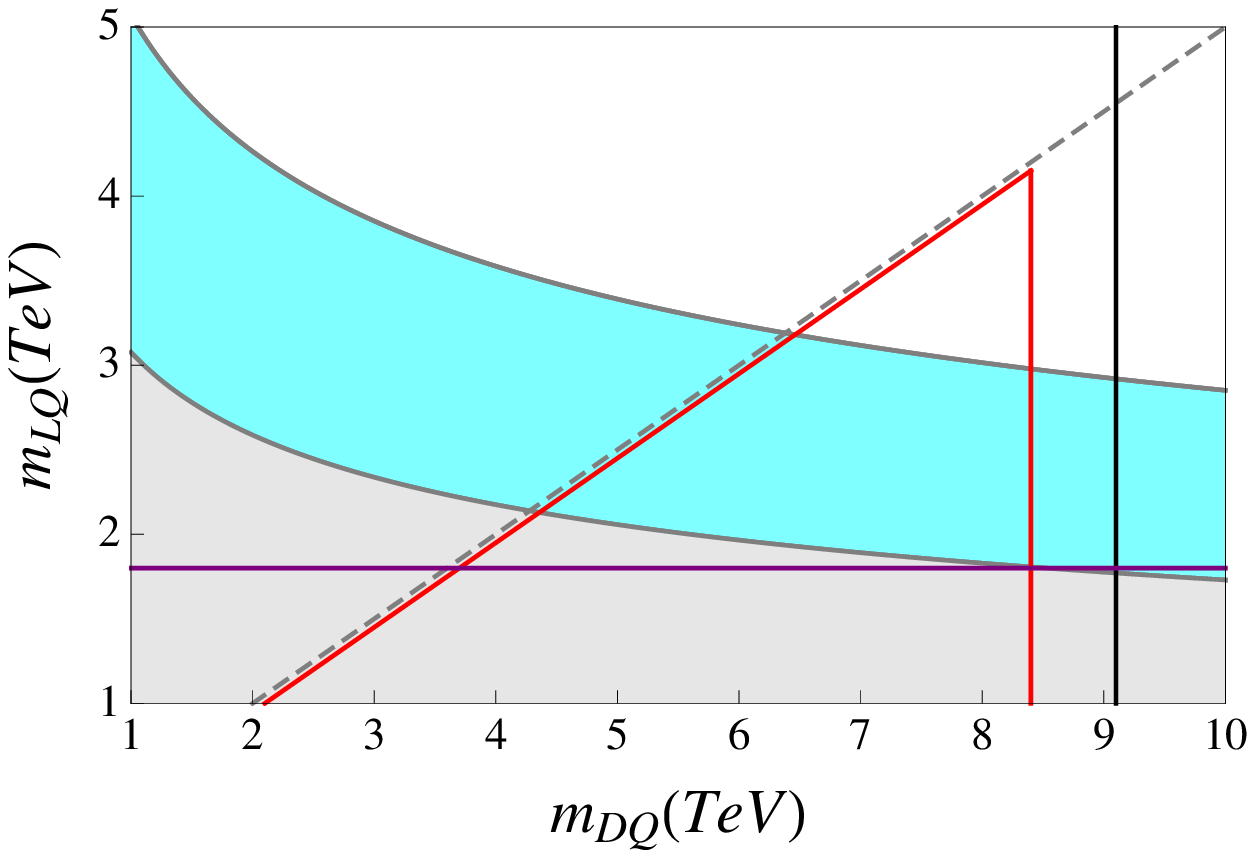}
\caption{\label{fig:DQ1} Expected future sensitivities for the LHC at
  $\sqrt{s} = 13 \ \text{TeV}$, ${\cal L} = 20/$fb (left) and ${\cal
    L} = 300/$fb (right), compared with current and future double beta
  decay experiments for the diquark model described in the Lagrangian
  eq. (\ref{eq:3-modif-1}). The vertical black line corresponds to future
  limits coming from dijet searches at the LHC, the horizontal purple
  line from leptoquark searches and the triangular red curve covers
  the region for like sign leptons plus two jets search. We use the
  parameters $g_1 = g_2 = g_L$ (bottom) and $g_1 = g_2 = 0.2$ (top).
  $\mu$ is taken as $\mu = \frac{m_{DQ}}{6}$ (bottom) and $\mu =
  m_{DQ}$ (top).  The gray region corresponds to the current lower
  limit for the $\znbb$ decay half-life of $^{136}$Xe, the blue one
  corresponds to the estimated future sensitivity of $T_{1/2} = 6
  \times 10^{27}$ ys of the nEXO proposal.  The dashed line marks the
  kinematic limit for the $lljj$ search, where $m_{DQ}=2\times
  m_{LQ}$.  For more details see text.}
\end{figure}

In Fig. \ref{fig:DQ1} we show a comparison between $\znbb$ decay and
dijet, LQ and dilepton plus jets searches at LHC in the plane $m_{DQ}$
vs $m_{LQ}$, for two fixed choice of $g_1=g_2 $ (bottom: $g_1=g_L$,
top: $g_1=0.2$) and two values for the accumulated luminosity: ${\cal
  L} = 20/$fb (left) and ${\cal L} = 300/$fb (right).  Here, $g_L$ is
the $SU(2)_L$ coupling. $\mu$ is chosen as $\mu = m_{DQ}/6$ (bottom)
and $\mu = m_{DQ}$ (top). The vertical black line corresponds to
future limits from dijet searches at the LHC, the horizontal purple
line is for leptoquark searches and the triangular red curve covers
the region probed by the $lljj$ search. The dashed line shows the
kinematic limit for the $lljj$ signal, where $m_{DQ}=2\times
m_{LQ}$. For masses $m_{DQ} < 2\times m_{LQ}$, one of the LQs goes
off-shell and the branchig ratio for the final state $lljj$ drops to
unmeasurably small values.

As the figs (\ref{fig:DQ1}) on the left show, LHC searches will
significantly constrain parameter regions of LNV models contributing
to $\znbb$ decay already with moderate luminosities. The $lljj$ signal
depends very sensitively on the choice of $\mu$, while the dijet
signal depends mostly on the value of $g_1$. Smaller values of $\mu$
reduce the branching ratio for the $lljj$ final state, reducing 
its reach. However, in this case the branching ratio for the dijet
final states increases, making the dijet search more powerful, as the
figure shows.  We stress again, that while dijet searches can be used
to exclude parameter regions of LNV models contributing to $\znbb$
decay, to establish a direct relation between $\znbb$ and LHC, a
positive result from the LNV search ($lljj$) at the LHC would be
necessary.

For ${\cal L} = 300/$fb, see fig. (\ref{fig:DQ1}) on the right, the
LHC can probe up to DQ masses of the order of $8-9$ TeV (for $g_1 \ge
0.2$). Whether dijet or LNV signal are more constraining depends on
the exact value of $\mu$ We have chosen the value of $\mu=m_{DQ}/6$,
because, as the figure on the bottom right shows, negative results
from LHC LQ and dijet searches would rule out partial $\znbb$ decay
half-lives in this model below the current experimental limit for $\mu
= m_{DQ}/6$, assuming $g_1=g_2=g_L$.  For $\mu \le m_{DQ}/50$ negative
searches from the LHC would rule out partial $\znbb$ decay half-lives
below the future bound of $T_{1/2} = 6 \times 10^{27}$ ys.

$\znbb$ decay depends on the mean of the couplings and masses, see
eq. (\ref{eq:eps}). Thus, in general LHC and $\znbb$ decay probe
complementary parts of parameter space.  This can also be seen 
in fig.  (\ref{fig:DQ1}): For large values of $\mu$ and/or large 
values of $g_1$ and $g_2$ there is always a region in parameter space 
for large values of the DQ mass, where double beta decay is more 
sensitive than the LHC.

\begin{figure}
\centering
\includegraphics[scale=0.61]{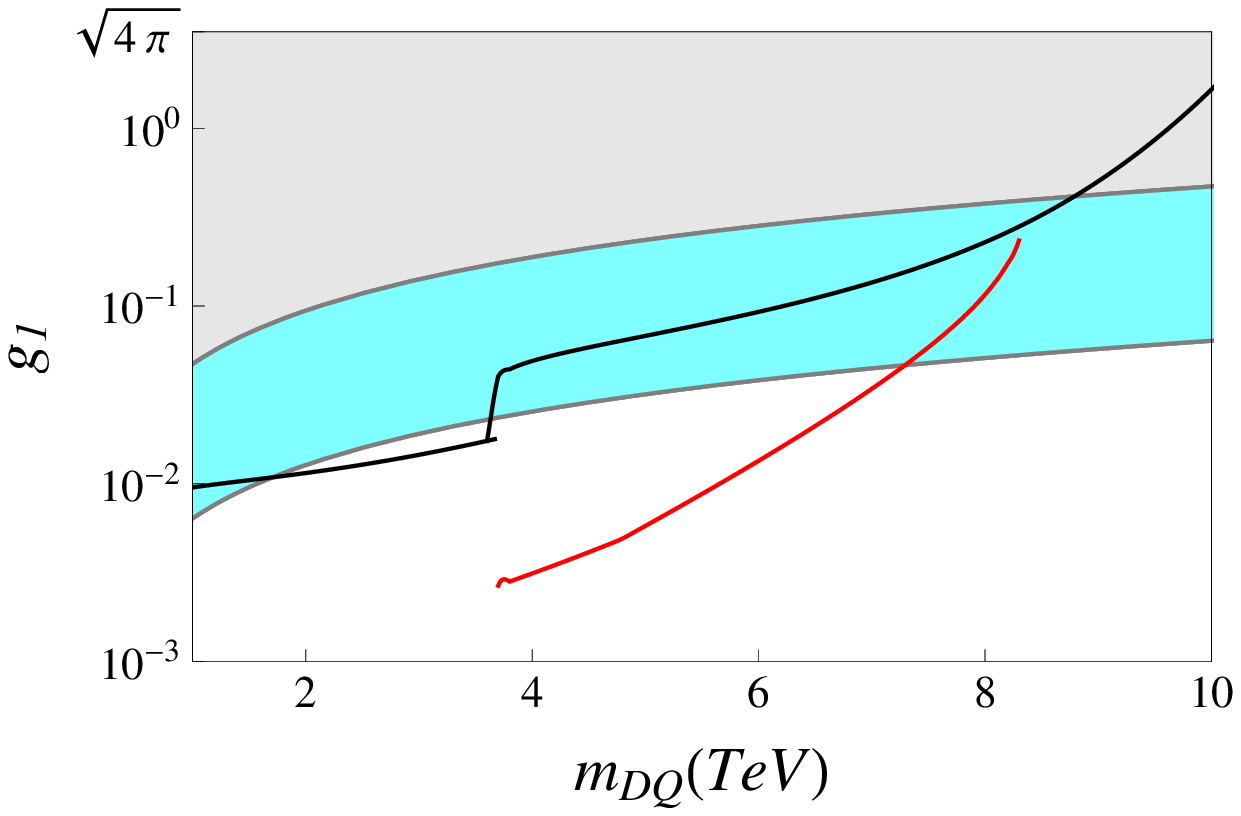}
\includegraphics[scale=0.61]{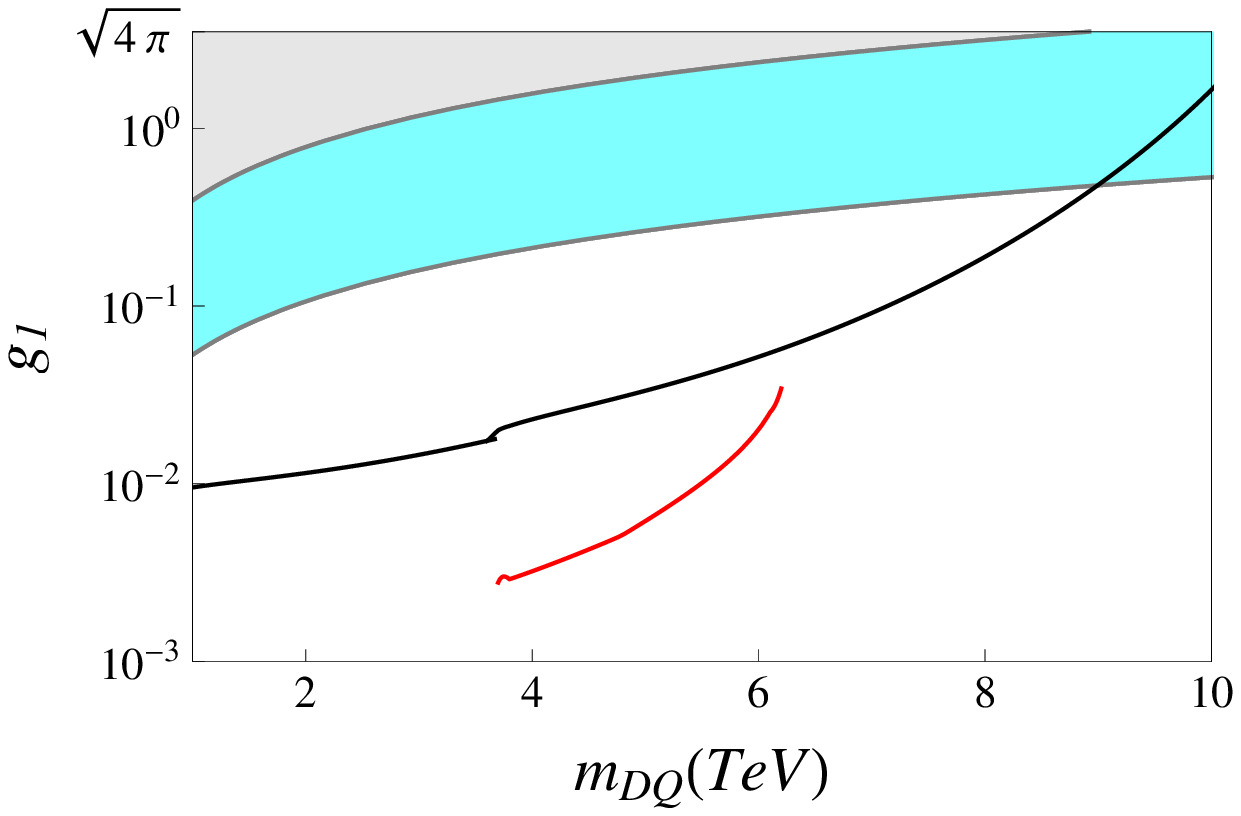}
\caption{\label{fig:DQ2} Regions in parameter space of the diquark
  model described in the Lagrangian (\ref{eq:3-modif-1}), which can be
  probed by dijet (black curves) and like sign leptons plus two jets
  (red curves) searches at LHC at $\sqrt{s} = 13 \ \text{TeV}$ and
  ${\cal L} = 300 \ \text{fb}^{-1}$. We use the parameters $ m_{LQ} =
  1.8$ TeV, $g_2 = g_L$, $\mu = \frac{m_{DQ}}{6} $ (left) and $\mu =
  \frac{m_{DQ}}{50} $ (right). The gray region is the current lower
  limit in $\znbb$ decay half-life, the blue one the estimated future
  sensitivity of $T_{1/2} = 6 \times 10^{27}$ ys. For more details see
  text.}
\end{figure}

In Fig. \ref{fig:DQ2} we show the comparison between the $\znbb$ decay
and dijet and dilepton plus jets searches at LHC in the plane
$g_1-m_{DQ}$. The LQ mass was chosen as $ m_{LQ} = 1.8$ TeV, roughly
the expected future bound from LHC. $g_2 = g_L$, $\mu =
\frac{m_{DQ}}{6} $ (left) and $\mu = \frac{m_{DQ}}{50} $ (right). 
Grey and blue regions show again the sensitivity of $\znbb$ decay 
current and future. The solid lines correspond to future
LHC limits from dijet (black curves) and dilepton plus jets (red
curves). The red curves start at $m_{DQ}=2\times m_{LQ}$ and 
stop at masses of the DQ, for which there are less than 5 
signal events expected in ${\cal L} = 300 \ \text{fb}^{-1}$. 

For these choices of parameters, dijet searches can probe larger
masses, but the $lljj$ search probes smaller values of the coupling
$g_1$. Again, for larger choices of $\mu$ the branching ratio for the
$lljj$ final state is larger and the $lljj$ search becomes more
sensitive. Negative results from the dijet searches would exclude 
large part of the parameter space explorable by future $\znbb$ decay 
experiments. However, for large values of $\mu$ there is always a 
corner of parameter space for large couplings and DQ masses, where 
$\znbb$ decay is more sensitive.

 \begin{figure}
\centering
\includegraphics[scale=0.61]{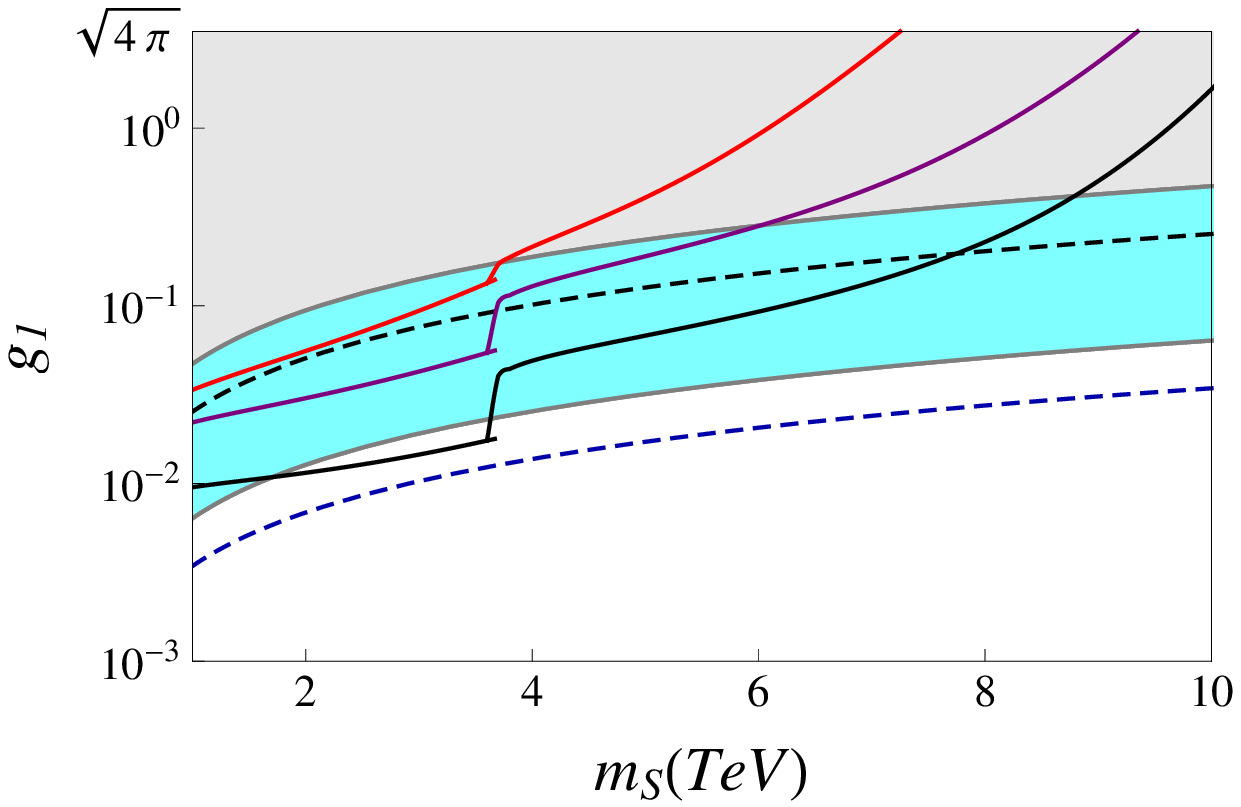}
\includegraphics[scale=0.61]{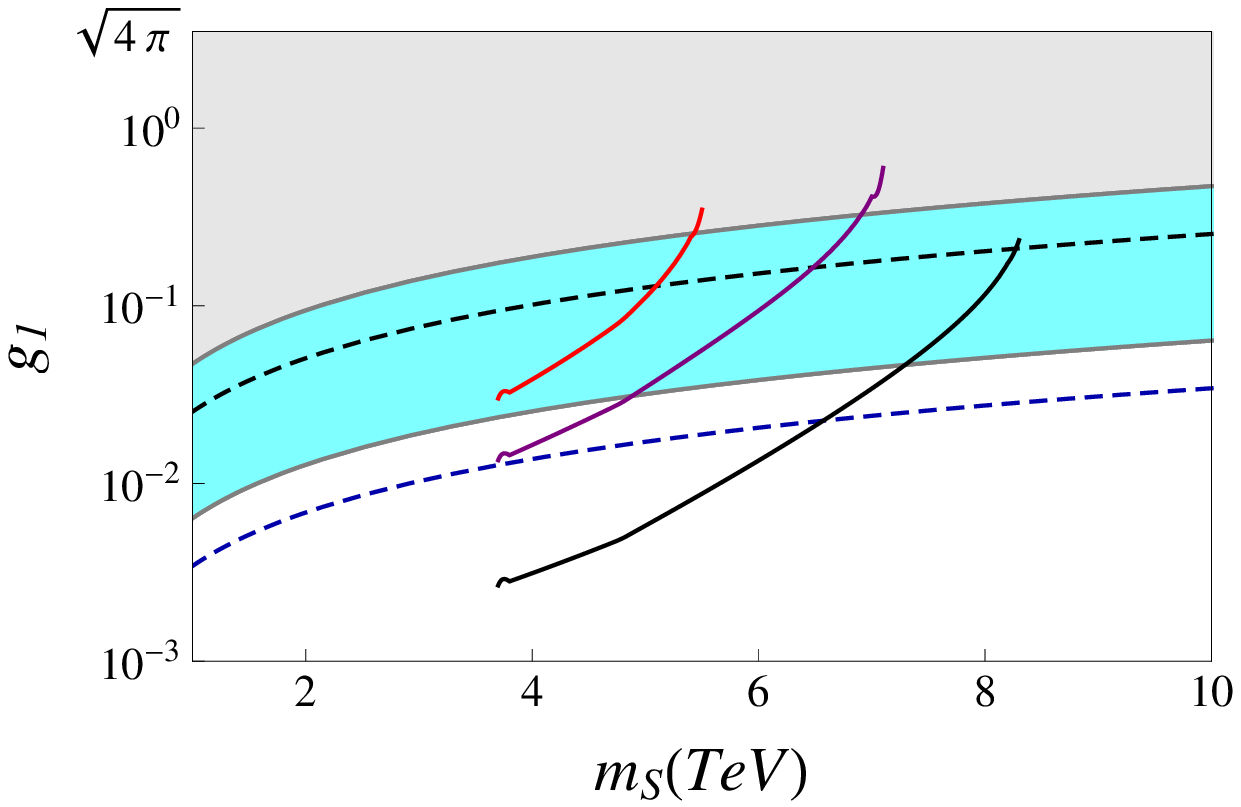}
\caption{\label{fig:DQ3} Future limits for the LHC at $\sqrt{s} = 13
  \ \text{TeV}$ and ${\cal L} = 300 \ \text{fb}^{-1}$ compared with
  current and future double beta decay experiments. The gray region is
  the current lower limit in $\znbb$ decay half-life whereas the blue
  region represents the parameter region accessible in near future
  $\znbb$ experiments. The colored lines shows sensitivity limits for
  the LHC for dijet (left) and dilepton plus jets (right) searches for
  production of three different scalar bosons $S_{+1}$(red),
  $S_{2/3}^{DQ}$ (purple) and $S_{4/3}^{DQ}$ (black). These limits
  were calculated using $g_2 = g_L$ and $m_{LQ} = 1.8$ TeV and $\mu
  =\frac{m_{DQ}}{6}$. For more details see text. }
\end{figure}

Finally in Fig. \ref{fig:DQ3} we plot a comparison of sensitivities of
$\znbb$ decay and the dilepton plus jets (Fig. \ref{fig:DQ3} right)
and dijet (Fig. \ref{fig:DQ3} left) searches at LHC for the three
different models discussed in section \ref{sec:decomp}: T-II-2 BL $\#$
11 (singly charged scalar), T-II-4 BL $\#$ 11 (triplet diquark) and
T-II-5 BL $\#$ 11 (singlet diquark). The double beta decay and LHC
limits were calculated using the parameters $\mu = m_{DQ}/6 $, $m_{LQ}
= 1.8$ TeV and $g_2 = g_L$.  The LHC is most sensitive for the case of
the triplet diquark model (T-II-2 BL $\#$ 11), black curve. This is
simply because the cross section of the resonance production of the scalar
diquark $S_{DQ}^{(4/3)}$ is larger than the one for the diquark
$S_{DQ}^{(-2/3)}$ (purple curve) and the singled charged scalar
$S^{1}$ (red curve). Fig. \ref{fig:DQ3} shows also current
and future limits from $0 \nu \beta \beta$ decay for the respective
models in consideration. The gray area is the currently excluded part
of parameter space from non observation of $^{136}$Xe decay with
$T_{1/2} > 1.1 \times 10^{26}$ yr and the blue one the estimated
future sensitivity, as before. The full lines are for the two 
diquark models (which have the same nuclear matrix elements, see 
above). The dashed lines are for the singly charged scalar model 
(T-II-2 BL $\#$ 11), which has a different nuclear matrix element, 
compare eqs (\ref{eq:MDQ}) and (\ref{eq:MS1}). LHC is least sensitive 
for the singly charged scalar case, the $S_{DQ}^{(-2/3)}$ is intermediate 
between the other two.

Finally, we briefly comment on other T-II models. As shown 
in table \ref{Tab:TopoII}, all T-II decompositions contain 
either a diquark or a charged scalar (in one case two different 
diquarks). The three example models, which we used in the numerical 
analysis, covers the cases with the largest and smallest cross 
sections at the LHC. It also covers the models with the largest 
and smallest matrix elements for the $\znbb$ decay. Thus, our 
sensitivity estimate for the future covers the extreme cases, 
both optimistic and pessimistic, and all other models should lie 
somewhere in between. 

In case of a discovery in the future at the LHC, one important
question to ask is, which of the different model possibilities is the
one realized in nature. As in the case of T-I  \cite{Helo:2013ika}, 
this might be achieved by investigating mass peaks in different 
variables and by the measurememt of the ``charge asymmetry'', 
i.e. the measurement of the number of events in $l^-l^-jj$ relative 
to  $l^+l^+jj$.

\section{Discussion and summary}
\label{sect:cncl}
 
We have discussed how future LNV and dijet searches at the LHC can be
used to constrain scalar short-range contributions to neutrinoless
double beta decay (topology-II diagrams). We have concentrated on
three LNV models, chosen from the full list of possible scalar
short-range contributions to $\znbb$ decay given in
\cite{Bonnet:2012kh}. Two of these models contribute to $\znbb$ decay
through short-range diagrams mediated by diquark scalars and one of
them by a singly charged scalar. For these models we have shown that
the future LNV and dijet searches at the LHC will provide stringent
constraints on the parameter space of the models, complementary to
$\znbb$ decay experiments. Except for small parts of the parameter
region of these LNV models, a $\znbb$ decay signal corresponding to a
half life in the range $T_{1/2} < 10^{27}$ ys should imply a positive
LNV or dijet signal at the LHC. On the other hand, the non-observation
of a positive signal at the LHC would rule out most of the parameter
region measurable in $\znbb$ decay. We note that, while we have
concentrated on three particular examples, similar constraints will
apply to any scalar short-range contributions to $\znbb$.

Finally, we mention that the observation of lepton number violation at
the LHC and/or in double beta decay will have important consequences
for high-scale models of leptogenesis
\cite{Deppisch:2013jxa,Deppisch:2015yqa}.

\bigskip
\bigskip
\centerline{\bf Acknowledgements}

\medskip
This work was supported by the Spanish grants FPA2014-58183-P,
Multidark CSD2009-00064 and SEV-2014-0398 (from the \textit{Ministerio
  de Economía y Competitividad}), as well as PROMETEOII/2014/084 (from
the \textit{Generalitat Valenciana}). J.C.H. is supported by Chile
grants Fondecyt No. 1161463, Conicyt ACT 1406 and  Basal FB0821. L. G. is supported by Conicyt (Chile) under Grant No. 21160645 and 
by the DGIP of the UTFSM. S. K. is supported by  FONDECYT (Chile) grant No. 1150792, CONICYT (Chile)  Ring ACT 1406 and Basal FB0821.

\end{document}